\documentclass{iau}

\usepackage{amsmath}
\usepackage{graphicx}
\usepackage{multirow}
\usepackage{multicol}
\usepackage[justification=raggedright,singlelinecheck=false]{caption}

\begin{document}

\lefttitle{Molina-Calzada et al.}
\righttitle{ALS IV: A massive census in the MCs}

\jnlPage{1}{7}
\jnlDoiYr{2025}
\doival{10.1017/xxxxx}

\aopheadtitle{Proceedings IAU Symposium}
\volno{402}
\editors{A. Wofford,  N. St-Louis, M. García \&  S. Simón-Díaz, eds.}

\title{Alma Luminous Star catalogue IV: A Massive star census in the Magellanic Clouds}

\author{J. A. Molina-Calzada$^{1, 2}$ and J. Maíz Apellániz$^{1}$}
\affiliation{$^1$Centro de Astrobiología, CSIC-INTA, Campus ESAC, Camino bajo del castillo s/n, E-28\,692, Villanueva de la Cañada, Madrid, Spain; email: jmolina@cab.inta-csic.es\\
$^2$Departamento de Física de la Tierra y Astrofísica, Universidad Complutense de Madrid (UCM), E-28\,040 Madrid, Spain}

\begin{abstract}

    The fourth part of the Alma Luminous Star catalogue (ALS IV) aims to create the most comprehensive sample of massive stars in the Magellanic Clouds (MCs). By combining \textit{Gaia} DR3 with Simbad and complementing this information with other photometric and spectroscopic catalogues, we select the massive stars in this region. To achieve this, we apply filters in photometry, combining different bands, as well as in variability and spectral types from the literature. With this approach, we will obtain one of the most complete samples of massive stars in the MCs, which can be used both to study the Clouds and the Magellanic Bridge, as well as the massive stars they contain.
    
\end{abstract}


\maketitle

\vspace{-0.2cm}
\section{Introduction}

The Alma Luminous Stars (ALS) catalogue aims to provide the most complete compilation of massive stars to date. It was first introduced by \citet{2003AJ....125.2531R} (ALS I), who collected $UBV\beta$ photometry for Milky Way massive stars. After the release of \textit{Gaia} DR2, a cross-match provided updated astrometric and photometric data (\citealt{2021MNRAS.504.2968P}; ALS II), allowing the sample to be refined with high completeness in the solar neighborhood. With \textit{Gaia} DR3, the catalogue reached about 14\,000 massive stars (\citealt{2025MNRAS.543...63P}; ALS III). The study is now being extended to the Magellanic Clouds (MCs) with ALS IV, which will provide the most extensive sample of massive stars in the Large Magellanic Cloud (LMC), Small Magellanic Cloud (SMC), and Magellanic Bridge (MB).

Massive stars are defined as those above 8~M$_\odot$ but converting that into observational characteristics is not trivial. Here we identify OB stars (not to be confused with all stars of spectral types O and B, \citealt{2026enap....2...43M}), A to M supergiants, Wolf–Rayet stars (WRs), and related types, all of them luminous objects in the Color–Magnitude Diagram (CMD).

\section{Data}

To build a full catalogue of massive stars in the MCs, we start with two key surveys. \textit{Gaia} DR3 \citep{2023A&A...674A...1G}, which provides the photometric and astrometric data, and Simbad \citep{2000A&AS..143....9W}, which collects data from the literature. These catalogues will be complemented with other supplementary data: $UBV$ photometry from \citet{2002AJ....123..855Z, 2004AJ....128.1606Z} and other surveys, and infrared data from VISTA \citep{2011A&A...527A.116C} and 2MASS \citep{2006AJ....131.1163S}, among others.

\section{Filtering process}

\begin{figure}[ht]
    \centering
    \includegraphics[width=0.289\textwidth]{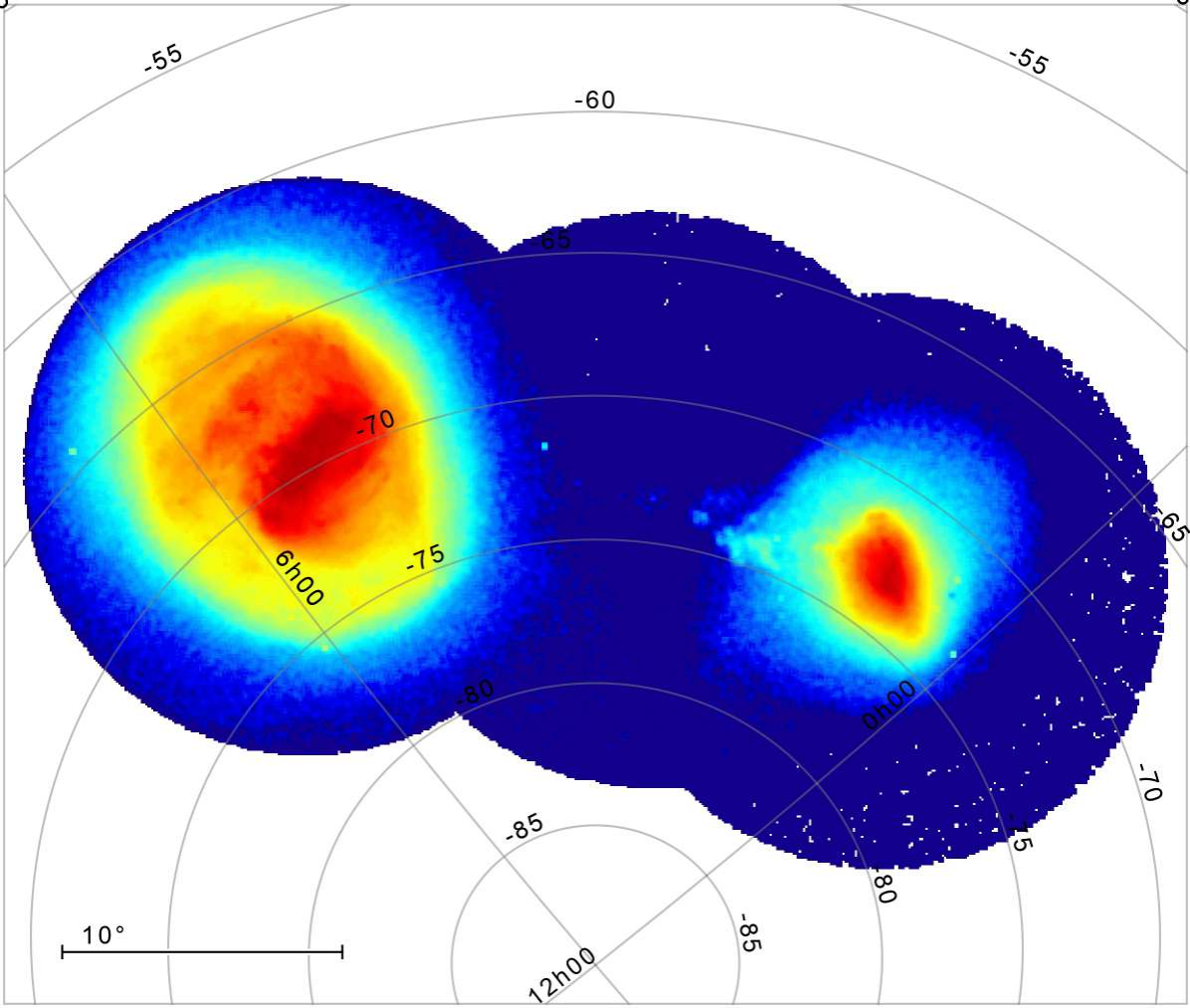}
    \hfill
    \includegraphics[width=0.294\textwidth]{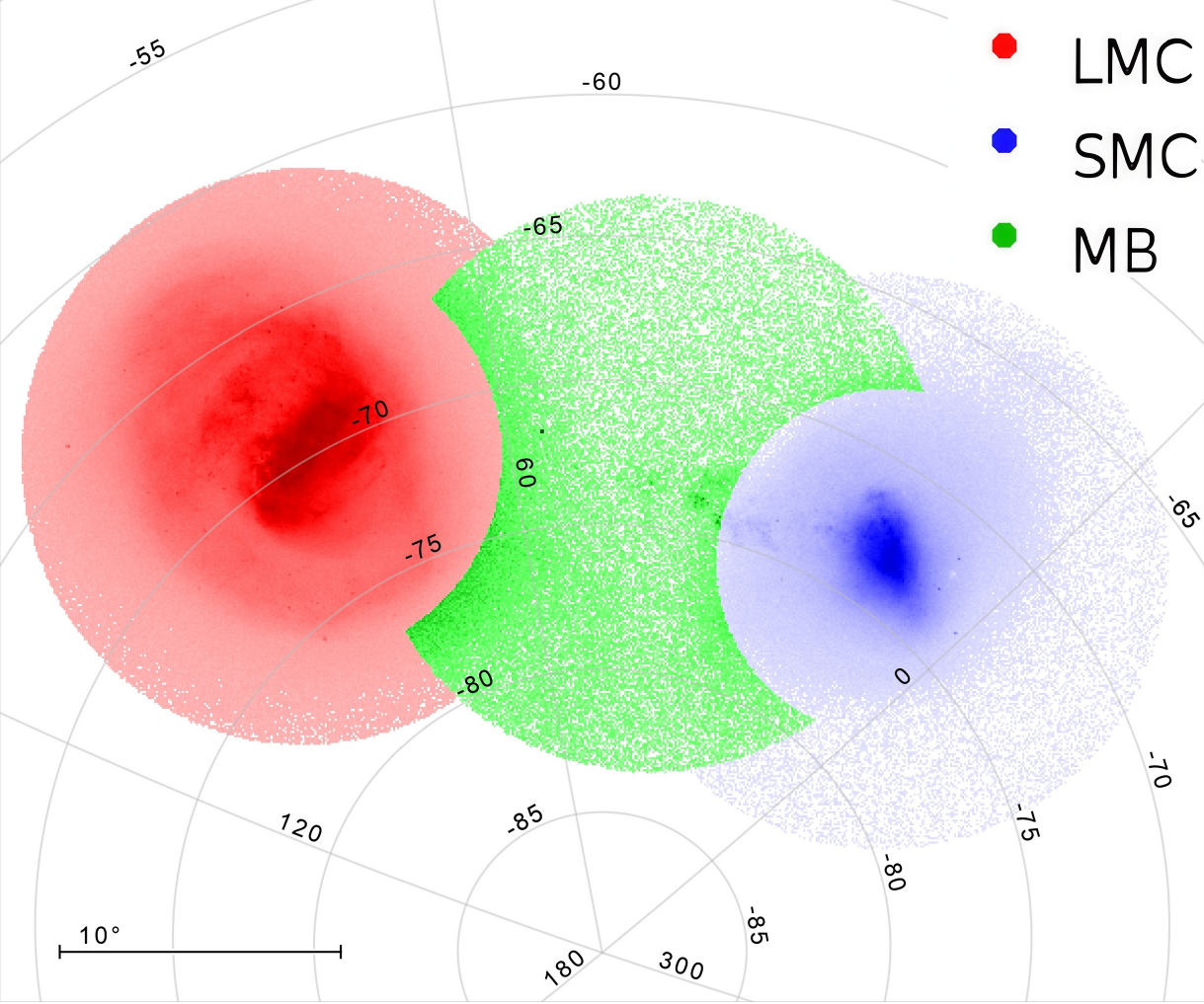}
    \hfill
    \includegraphics[width=0.322\textwidth]{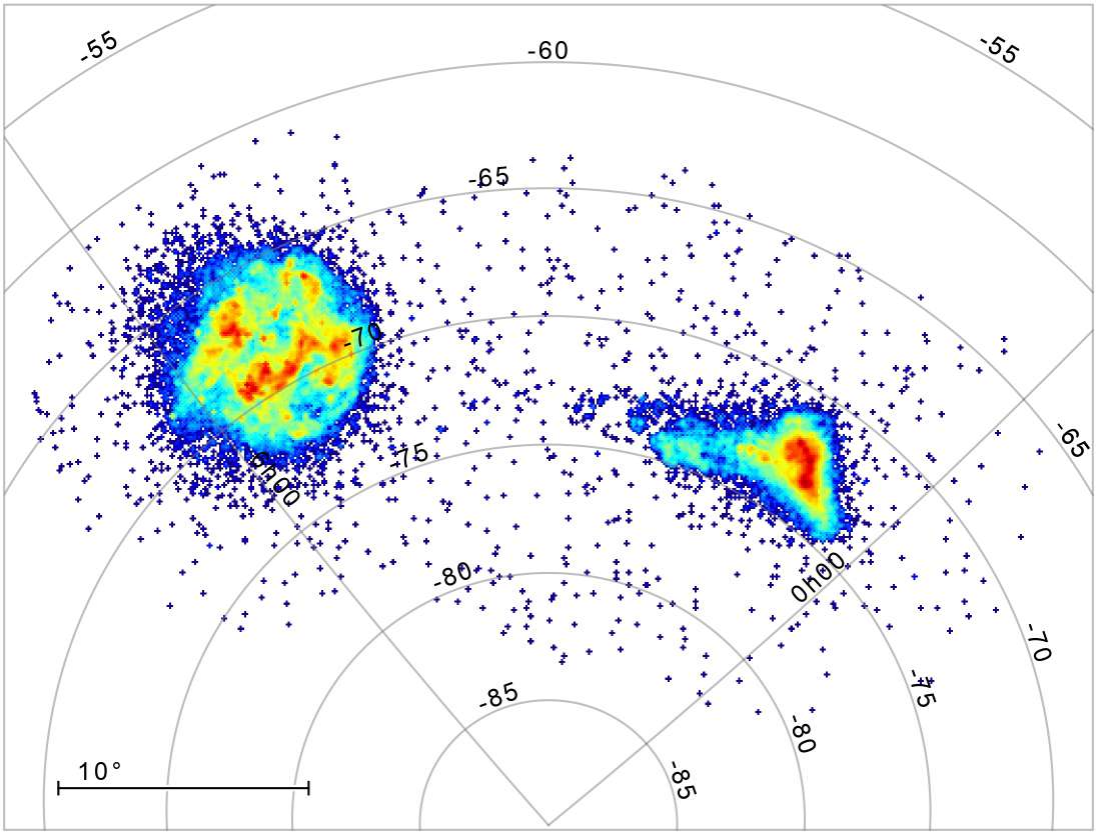}
    \caption{Sky map of the \textit{Gaia} sample for the MCs (left), the defined regions (center; the intensity scales differ for each color), and the filtered LG, SG, and BG samples (right).}
    \label{fig: sky_plots}
\end{figure}

We downloaded \textit{Gaia} DR3 and Simbad data for the MCs. Figure \ref{fig: sky_plots} shows a sky map of the \textit{Gaia} sources and the defined regions for the LMC, SMC, and MB.

\begin{figure}[ht]
\begin{center}
    \vspace{-3mm}
    \includegraphics[scale=0.37]{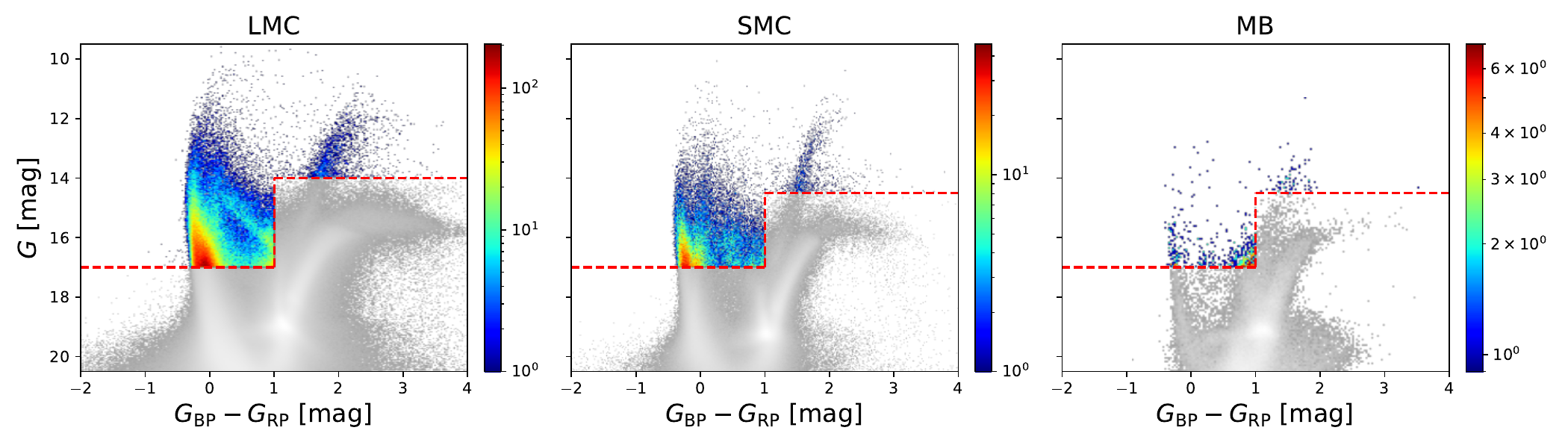} 
    \caption{Photometric cut applied to the CMD of \textit{Gaia} DR3, after astrometric filtering, for the LMC (left), SMC (center), and MB (right).}
    \label{fig: photometric_cut}
\end{center}
\end{figure}

\begin{figure}[ht]
\begin{center}
    \vspace{-10mm}
    \includegraphics[scale=0.37]{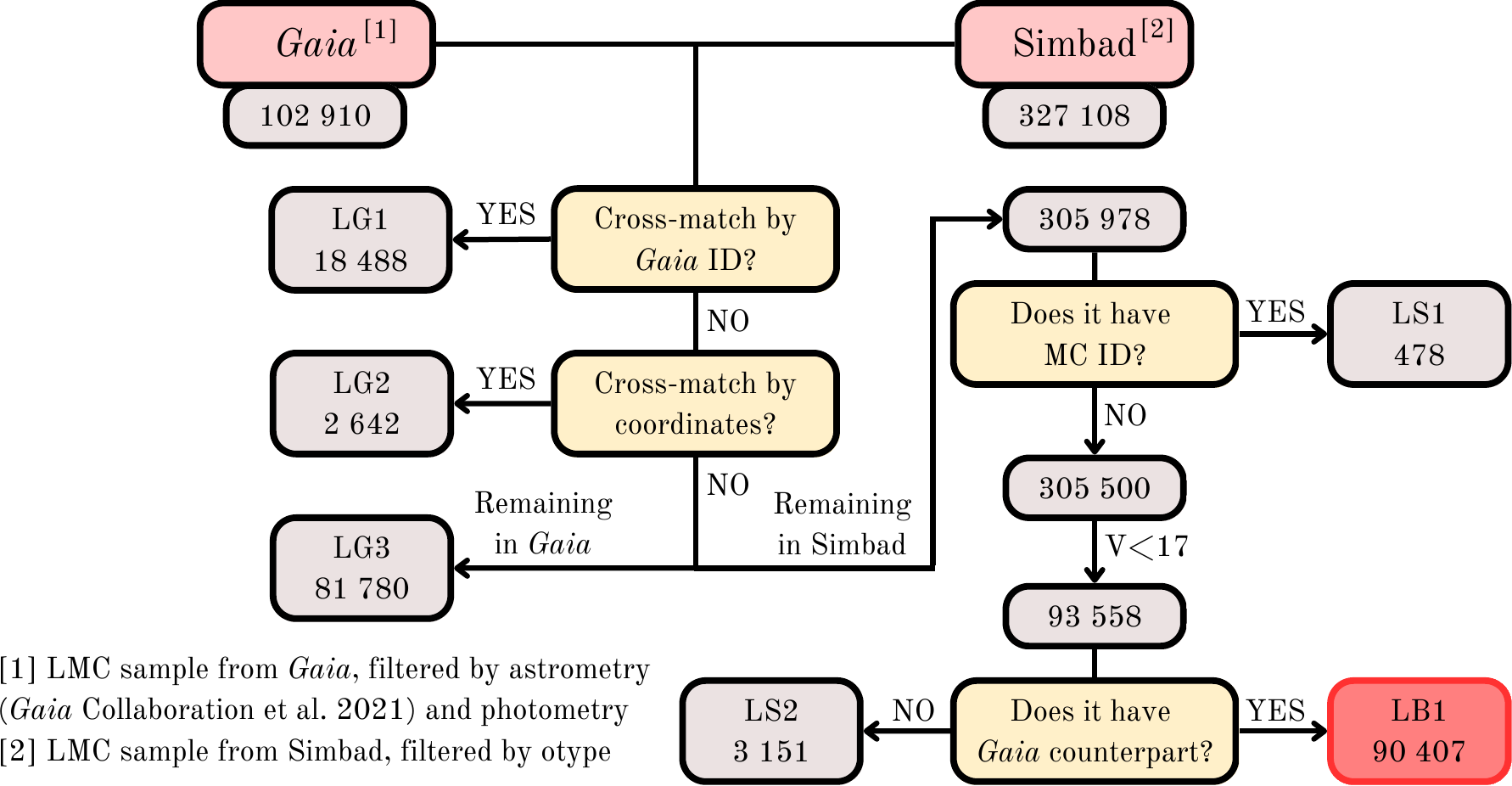} 
    \caption{Flow chart of the cross-match between \textit{Gaia} DR3 (filtered astrometrically and photometrically) and Simbad (filtered by stellar objects) for the LMC.}
    \label{fig: flow_chart_LMC}
\end{center}
\end{figure}

\vspace{-4mm}
The \textit{Gaia} sample is first filtered astrometrically, following the procedure described in \citet{2021A&A...649A...7G}, and then photometrically, by selecting the upper part of the CMD (see Figure \ref{fig: photometric_cut}). The Simbad sample was filtered by object type, keeping only stellar-type sources. Then, a cross-match is done between these two resulting samples. Figure \ref{fig: flow_chart_LMC} shows the results of the cross-match performed for the case of the LMC.

\begin{figure}[ht]
\begin{center}
    \includegraphics[scale=0.37]{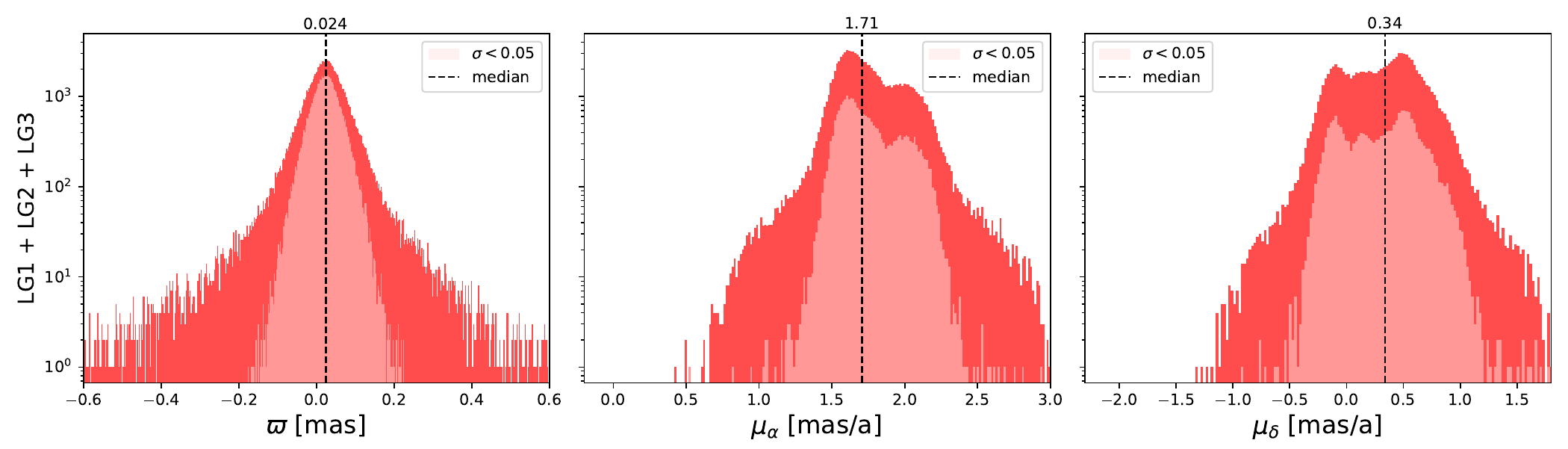} 
    \caption{Parallax (left) and proper motion (center and right) distributions (log scale) for the combined LG sample. Light shading: low uncertainties. Dashed line: median value. The structures seen in the proper motions reflect their internal motions.}
    \label{fig: hist_w_pmG123}
    \vspace{-5mm}
\end{center}
\end{figure}

\vspace{-8mm}
After completing the cross-match, we obtain the LG, SG, and BG samples for the LMC, SMC, and MB, respectively. Here, for brevity, we focus on the LMC, where the LG sample (LG1 + LG2 + LG3) corresponds to matches by identifier, by coordinates, and the remaining \textit{Gaia} sources. Figure \ref{fig: hist_w_pmG123} shows the parallax and proper motion distributions for the LG sample, showing its internal dynamical behavior.

\vspace{-10mm}
These final LG, SG, and BG samples still need cleaning, as the lower part of our photometric cut includes intermediate-mass stars. While one could apply an upper cut at the 8$M_\odot$ evolutionary track, extending the selection downward allows us to recover extinguished massive stars. To remove the remaining intermediate-mass objects, we use additional catalogues: $UBV$ photometry from \citet{2002AJ....123..855Z, 2004AJ....128.1606Z} to trace the temperature scale to exclude later-type stars, variability data from \citet{2023A&A...677A.137M} to remove non-massive Cepheids, or infrared surveys such as VISTA and 2MASS to identify extinguished stars, among others.

For the remaining objects that are only in Simbad, the process is more complex. First, we retain all sources with Cloud-specific identifiers (e.g., RMC, AzV, etc), yielding the LS1, SS1, and BS1 samples. A flux cut is then applied to keep only the brightest sources, and those with a \textit{Gaia} counterpart are excluded, since \textit{Gaia} objects in the cross-match were already selected from the Clouds. This produces the final LS2, SS2, and BS2 samples, which still require a bibliographic check to confirm the nature of the objects.

\section{Conclusions and Future Work}

The creation of a catalogue of massive stars in the MCs is a significant contribution to the scientific community. Even with preliminary filtering, structures of interest are already visible in the sky plots and in the parallax and proper motion distributions, revealing the internal dynamics of the LMC through distinct peaks.

To complete the study, we follow two complementary paths: for LG, SG, and BG samples, data from other photometric catalogues will be combined to remove intermediate-mass stars, while for LS, SS, and BS samples, the literature spectral classifications will be checked. These procedures will yield a complete sample of massive stars, enabling detailed studies of the MCs.

\end{document}